# Microwave photonic transversal filters based on microcombs with feedback control

*David J. Moss*

**Abstract— Feedback control plays a crucial role in improving system accuracy and stability for a variety of scientific and engineering applications. Here, we theoretically and experimentally investigate the implementation of feedback control in microwave photonic (MWP) transversal filter systems based on optical microcomb sources, which offer advantages in achieving highly reconfigurable processing functions without requiring changes to hardware. We propose four different feedback control methods including (1) one-stage spectral power reshaping, (2) one-stage impulse response reshaping, (3) two-stage spectral power reshaping, and (4) two-stage synergic spectral power reshaping and impulse response reshaping. We experimentally implement these feedback control methods and compare their performance. The results show that the feedback control can significantly improve not only the accuracy of comb line shaping as well as temporal signal processing and spectral filtering, but also the system's long-term stability. Finally, we discuss the current limitations and future prospects for optimizing feedback control in microcomb-based MWP transversal filter systems implemented by both discrete components and integrated chips. Our results provide a comprehensive guide for the implementation of feedback control in microcomb-based MWP filter systems in order to improve their performance for practical applications.**

## I. INTRODUCTION

**M**icrowave transversal filter systems with highly reconfigurable transfer functions have been widely used for achieving a variety of spectral filtering and signal processing functions [5-7]. Microwave photonic (MWP) transversal filter systems, which realize conventional microwave transversal filter systems based on photonic technologies, can provide processing bandwidths far beyond their electrical counterparts, which are limited by the electrical bandwidth bottleneck [8, 9]. They can also offer other attractive advantages, such as low loss when processing high-bandwidth signals, strong immunity to electromagnetic interference, and wide-band tunability [1, 8].

In MWP transversal filter systems, in order to achieve a high processing accuracy, a large number of optical carriers are required as discrete taps to sample the microwave signals to be processed. Conventional multi-wavelength sources such as discrete laser arrays [10-12] and fibre Bragg grating arrays [13-15] have been utilized to provide the discrete taps. Nevertheless, there exists a significant limitation in the available tap numbers (typically < 10) since their system size and complexity greatly increase with the tap number. In contrast, optical microcombs can simultaneously generate a large number of separated wavelengths based on micro-scale resonators with compact device footprint [6, 16-20], which makes them attractive for serving as multi-wavelength sources in the MWP transversal filter systems. Compared to laser frequency combs generated by electro-optic (EO) modulation [21-23] or mode-locked fiber lasers [24, 25], the large comb spacings of optical microcombs also yield large operation bandwidths for the MWP transversal filter systems by providing wide Nyquist bands between adjacent wavelength channels. Recently, a diverse range of functions have been realized based on microcomb-based MWP transversal filter systems implemented by either discrete components or integrated chip, such as differentiation [3, 26], integration [2], Hilbert transform [27, 28], arbitrary waveform generation [29, 30], filtering [17, 31], image processing [32], and neuromorphic computing [1, 4, 33, 34].

David J. Moss is with the Optical Sciences Center, Swinburne University of Technology, Hawthorn, VIC 3122, Australia.

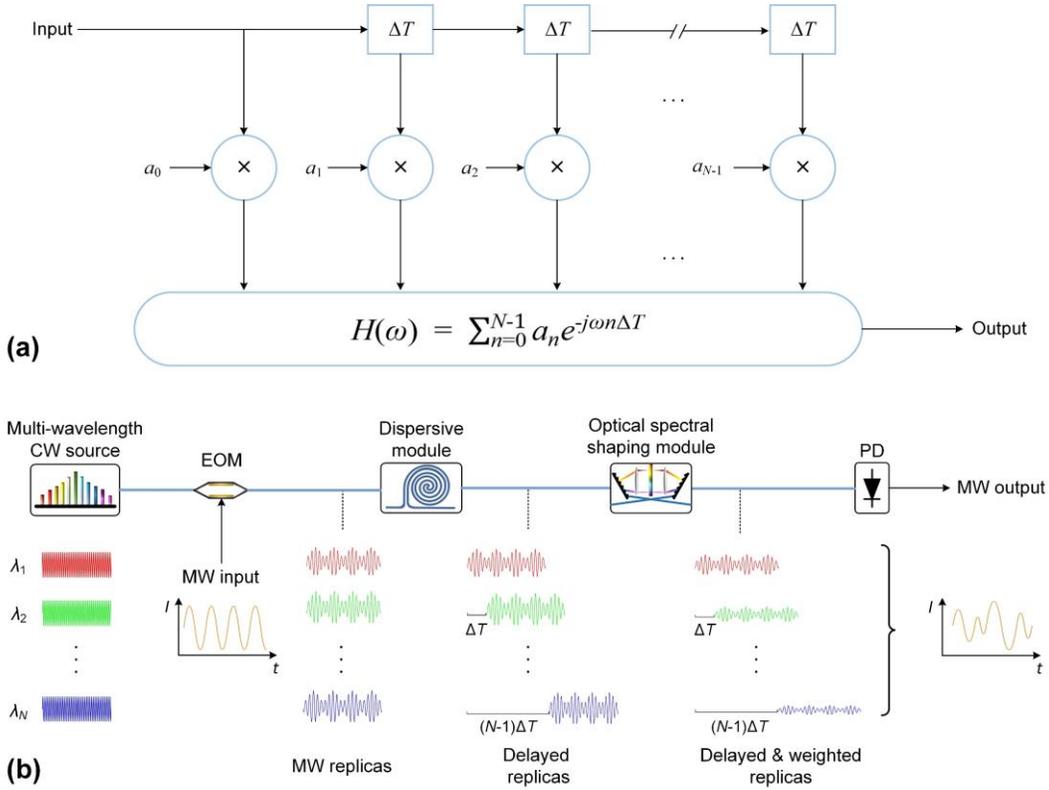

**Fig. 1.** Transversal filter systems. (a) General diagram of a microwave transversal filter system. (b) General diagram and processing flow of a microwave photonic (MWP) transversal filter system with a multi-wavelength continuous-wave (CW) source. EOM: electro-optic modulator. PD: photo detector. MW: microwave.

Feedback control techniques are of fundamental importance in many applications owing to their ability to enhance system stability and accuracy. They enable dynamic adjustments based on real-time measurements, ensuring that systems achieve desired response and maintain stable working states in the presence of disturbances or uncertainties. In MWP transversal filter systems with optical microcomb sources, the errors are mainly induced by imperfect response of experimental components [35-37], which can also be reduced by introducing feedback control. Here, we investigate feedback control in microcomb-based MWP transversal filter systems. We propose four different feedback control methods, including one-stage spectral power reshaping, one-stage impulse response reshaping, two-stage spectral power reshaping, and two-stage synergic spectral power reshaping and impulse response reshaping. We experimentally demonstrate feedback control based on these methods, and provide detailed comparison of their performance in improving the system accuracy with respect to comb line shaping, temporal signal processing, spectral filtering, as well as in enhancing the system stability in working for long times. Finally, we discuss the current challenges and future prospects for further improving the performance of feedback control in the systems implemented by both discrete components and integrated chips. These results provide valuable insights for the implementation of feedback control in microcomb-based MWP filter systems, facilitating improvements in their performance for practical applications.

## II. MICROCOMB-BASED MICROWAVE PHOTONIC TRANSVERSAL FILTER SYSTEMS

Transversal filter systems perform filtering functions when signals propagate in delay media, where different delayed signal replicas are tapped, weighted, and then summed to generate filter outputs. The system can be implemented by different configurations that provide delay elements, discrete taps, and a mechanism for weighting as well as summing the weighted replicas [38]. **Fig. 1(a)** shows the general diagram of a microwave transversal filter system. As the input signal propagates through the filter system, it is delayed by a delay line consisting of a series of delay elements, each with a time delay of $\Delta T$. After propagation through each delay element along the delay line, the delayed signal in each channel is weighted according to the designed tap coefficient. The delayed and weighted signals for different channels are then summed to produce the ultimate output. By adjusting the tap coefficients, different spectral responses can be achieved. In addition to spectral filtering, transversal filter systems can be used for temporal signal processing [39], where impulse responses for different processing functions can be realized via design of the inverse Fourier transform of the spectral responses [40].

**Fig. 1(b)** shows the general diagram and processing flow of a MWP transversal filter system with a multi-wavelength continuous-wave (CW) source. An input microwave signal is multicast onto different wavelength channels of the multi-wavelength source by using an electro-optic modulator (EOM). After the EO modulation, the microwave replicas are delayed by a dispersive module, which creates a time delay between adjacent channels. The delayed replicas then go through an optical spectral shaping module to get weights based on the designed tap coefficients. Finally, the delayed and weighted replicas are summed after photo detection and converted into an output microwave signal.

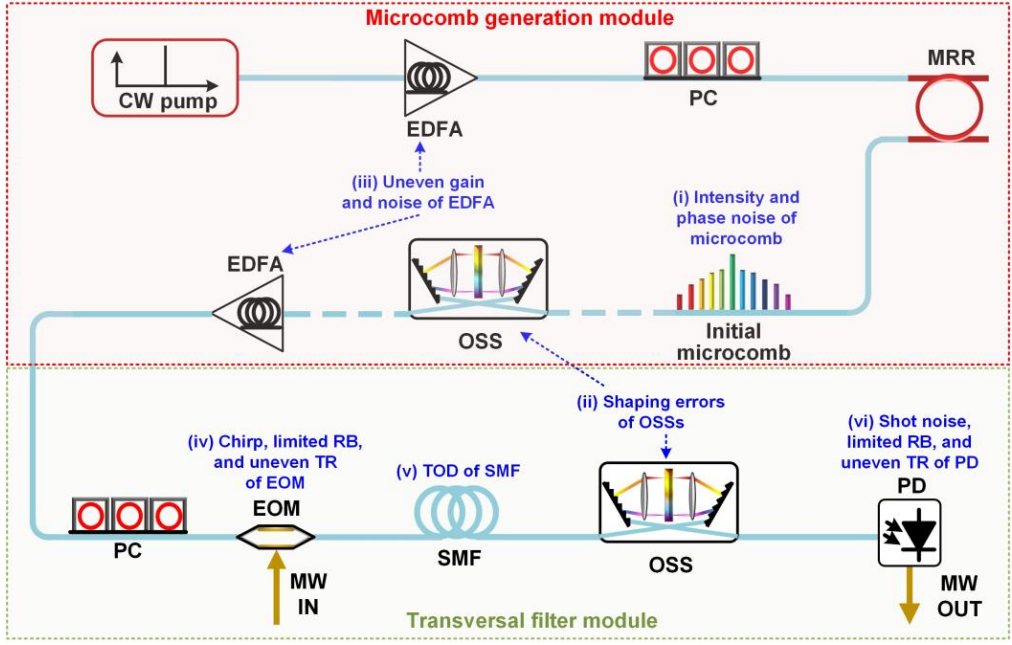

**Fig. 2.** Experimental schematic of a microcomb-based MWP transversal filter system together with experimental error sources that induce processing errors. CW pump: continuous-wave pump. EDFA: erbium-doped fibre amplifier. PC: polarization controller. MRR: microring resonator. OSS: optical spectral shaper. EOM: electro-optic modulator. MW: microwave. SMF: single-mode fibre. PD: photodetector. RB: response bandwidth. TR: transmission response. TOD: third-order dispersion.

The spectral transfer function of the MWP transversal filter system can be expressed as [41, 42]

$$H(\omega) = \sum_{n=0}^{N-1} a_n e^{-j\omega n \Delta T}, \qquad (1)$$

where $\omega$ is the angular frequency, $N$ is the number of taps, $a_n$ ($n$ = 0, 1, 2, …, $N$-1) is the tap weight of the $n^{th}$ tap, and $\Delta T$ is the time delay between adjacent taps. Note that the spectral transfer function in **Eq. (1)** is consistent with the spectral response of the microwave transversal filter system in **Fig. 1(a)**, although the system is implemented using MWP technologies. In contrast to the use of multiple electrical delay elements, attenuators, and accumulators in a microwave transversal filter system, a MWP transversal filter system employs only an optical delay module, an optical shaping module, and a PD to realize the delay, weighting, and sum functions, respectively. This makes MWP transversal filter systems provide an attractive advantage in achieving a low system complexity compared to microwave transversal filter systems, particularly given that the processing performance of a transversal filter system improves for an increased tap number [26].

**Fig. 2** shows a microcomb-based MWP transversal filter system, which consists of a microcomb generation module and a transversal filter module. In the microcomb generation module, a CW light is amplified by an erbium-doped fibre amplifier (EDFA) and used to pump a nonlinear microring resonator (MRR) with a high quality (Q) factor. A polarization controller (PC) is employed to adjust the polarization of the light fed into the MRR. The generated optical microcomb from the MRR, which serves as the multi-wavelength CW source in **Fig. 1(b)**, is sent to the subsequent transversal filter module. In the transversal filter module, a spool of single-mode fibre (SMF) is employed as the dispersive module. In principle, only one OSS is needed to shape the microcomb to achieve the target tap coefficients. Nevertheless, for practical systems, particularly when the initial optical microcomb generated by the MRR exhibits significant variations in power among its comb lines, two OSSs can be employed to improve the shaping accuracy and facilitate feedback control. As illustrated in **Fig. 2**, the first OSS after the MRR can be used to flatten the initial optical microcomb, thus leading to higher signal-to-noise ratios of the comb lines as well as improved loss control range for the second OSS. The second one in the transversal filter module is employed to shape the comb lines according to the designed tap coefficients.

The microcomb-based MWP transversal filter system in **Fig. 2** is essentially equivalent to the digital signal processing (DSP) filter in **Fig. 1(a)** but implemented by photonic hardware, which can not only maintain the high processing speed of photonic processing but also enable improved processing accuracy than optical analogue processing based on passive optical filters [41]. By programming the optical spectral shaping module to apply different tap coefficients, the same system setup can perform different functions without any changes of the hardware.

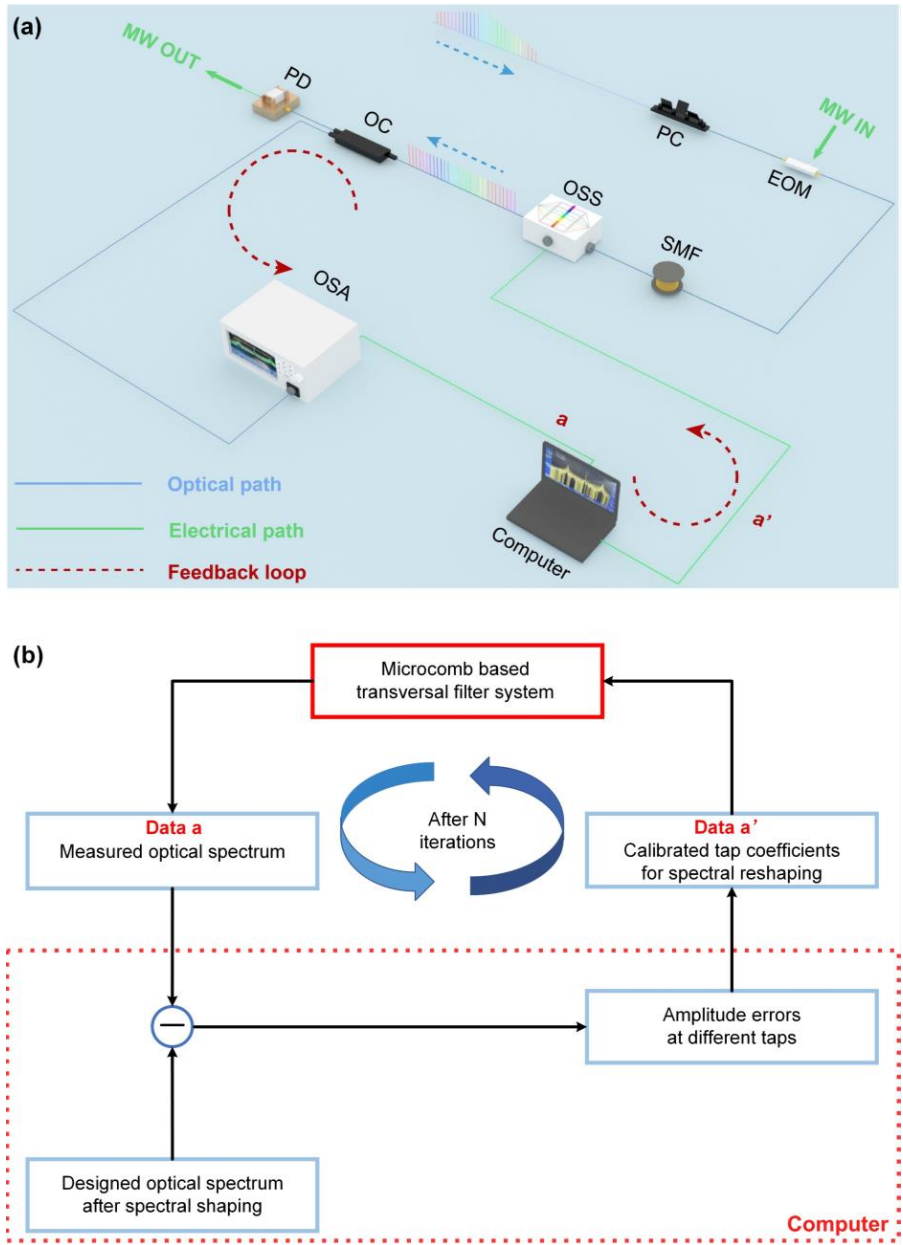

**Fig. 3.** Feedback control based on one-stage spectral power reshaping in a microcomb-based MWP transversal filter system. (a) Schematic of system setup. PC: polarization controller. EOM: electro-optic modulator. SMF: single-mode fiber. OSS: optical spectral shaper. PD: photo detector. OC: optical coupler. OSA: optical spectrum analyzer. (b) Flowchart of spectral power reshaping process in the feedback control loop.

For microcomb-based MWP transversal filter systems used for spectral filtering or signal processing, the filtering or processing errors are induced by both theoretical limitations and imperfect response of the experimental components in **Fig. 2**. When the tap number $N$ is high enough (*e.g.*, > 40, as in the case of optical microcombs), the accuracy is mainly constrained by the experimental factors [35]. As labelled in **Fig. 2**, the imperfect response of the experimental components is induced by several factors including (i) intensity and phase noise of optical microcomb, (ii) shaping errors of the OSSs, (iii) uneven gain and noise of the EDFA, (iv) chirp, limited response bandwidth (RB), and uneven transmission response (TR) of the EOM, (v) third-order dispersion (TOD) of the SMF, and (vi) shot noise, limited RB, and uneven TR of the PD.

The factors (i) – (vi) cause imperfect amplitude or phase response by introducing errors to the tap coefficients (*i.e.*, $a_n$ in **Eq. (1)**, $n = 0, 1, 2, …, N-1$) or time delay between adjacent taps (*i.e.*, $\Delta T$ in **Eq. (1)**). Among the various factors, the rapid fluctuations in the intensity and phase induced by optical microcomb and PD, typically exceeding 1 GHz, can be effectively alleviated by using advanced mode-locking technologies and highly sensitive PDs [16, 43]. With the exception of these two fast-changing error sources, the majority of error sources are relatively stable or change much more gradually, making them amenable to compensation through the implementation of feedback control aimed at adjusting the tap coefficients and time delays to match the designed values. In the following sections, the feedback control in microcomb-based MWP transversal filter systems is introduced and discussed in detail. In **Section III**, we introduce the principle of various feedback control methods. In **Section IV**, we experimentally implement feedback control based on different methods and compare their performance. In **Section V**, we discuss the current challenges and future prospects.

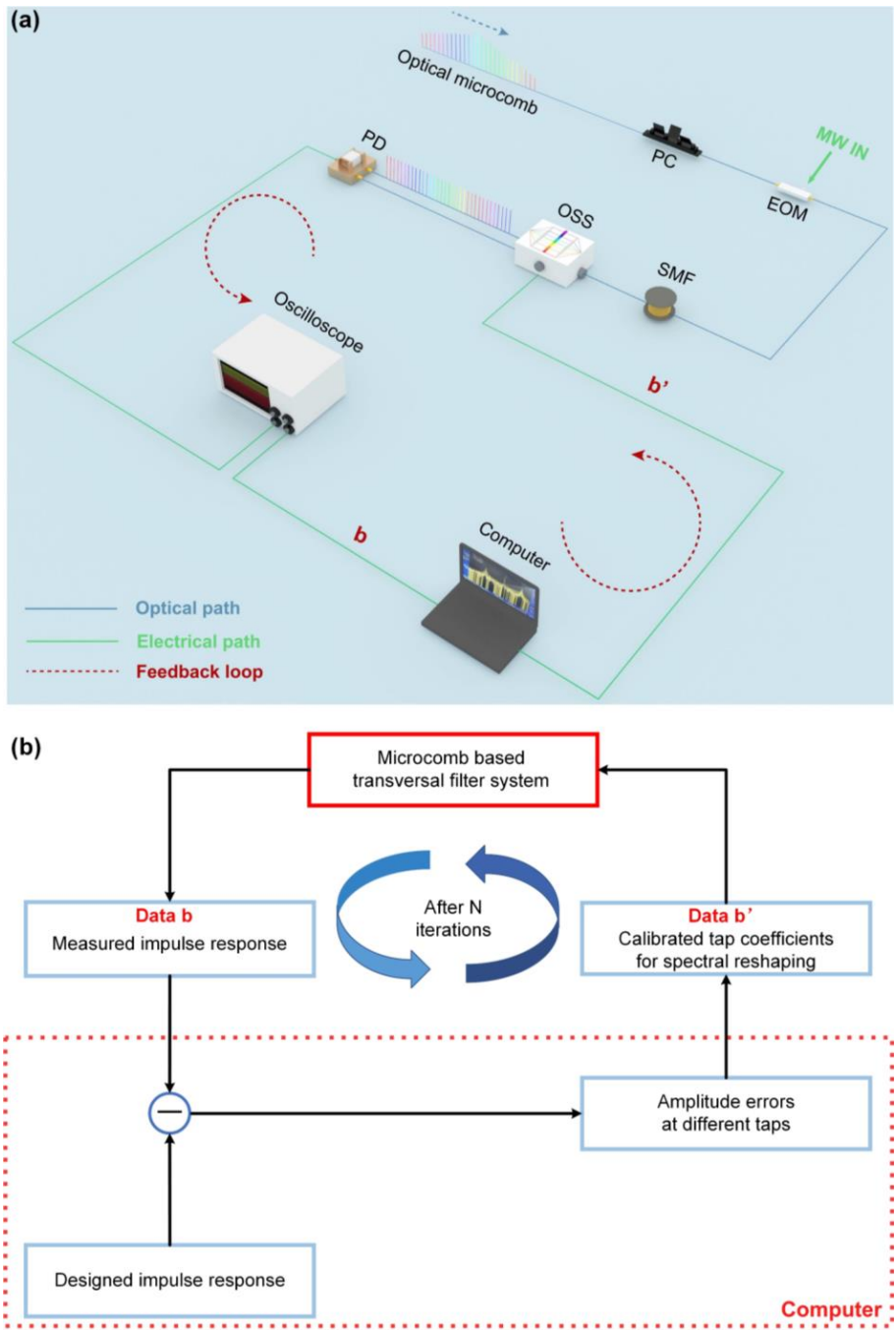

**Fig. 4.** Feedback control based on one-stage impulse response reshaping in a microcomb-based MWP transversal filter system. (a) Schematic of system setup. PC: polarization controller. EOM: electro-optic modulator. SMF: single-mode fiber. OSS: optical spectral shaper. PD: photo detector. (b) Flowchart of impulse response reshaping process in the feedback control loop.

## III. FEEDBACK CONTROL METHODS FOR MICROCOMB-BASED MICROWAVE PHOTONIC TRANSVERSAL FILTER SYSTEMS

In this section, we introduce various feedback control methods for microcomb-based MWP transversal filter systems. Depending on the number of feedback loops, the different methods are categorized as one-stage and two-stage feedback control, which are discussed in subsections A and B, respectively. Following these, comparison between different feedback control methods is provided in subsection C.

### A. One-stage feedback control

There are two typical methods to realize one-stage feedback control in microcomb-based MWP transversal filter systems, namely, spectral power reshaping and impulse response reshaping. The former is used to calibrate the spectral intensity of comb lines according to the designed tap coefficients, whereas the latter aims to calibrate the temporal impulse response of the transversal filter system according to the ideal impulse response of a specific processing function.

**Fig. 3(a)** illustrates the schematic diagram for feedback control based on spectral power reshaping. The optical microcomb generated by the microcomb generation module (not shown in **Fig. 3(a)**) features non-uniform power distributions [31] and enters the transversal filter module. Within this module, an OSS is inserted before the PD, with the purpose of performing spectral shaping to achieve the desired tap coefficients. To realize feedback control for calibration, the shaped optical spectrum of the output from the OSS is recorded by an optical spectrum analyzer (OSA), which is sent to a computer to compare with the ideal tap coefficients and then generate the calibrated tap coefficients that are fed into the OSS for spectral power reshaping.

**Figure 3(b)** shows the flowchart of the spectral reshaping process in the feedback control loop at each iteration, where $a$ and $a'$ are the data for the measured comb spectrum and the reshaped comb spectrum after calibration, respectively. When the calibrated data $a'$ are applied to the OSS, one iteration is finished. After that, the reshaped comb lines after calibration will be treated as the new data $a$ and employed as the new input of the feedback control loop for a subsequent iteration. The system normally requires multiple iterations to effectively reduce the comb power shaping errors, which can improve the signal-to-noise ratio and ensure more uniform link gain across the different wavelength channels.

**Fig. 4(a)** shows the schematic diagram for feedback control based on impulse response reshaping. A microwave signal is used as an input signal to test the impulse response of the MWP transversal filter system. The reshaping is performed channel by channel with the same input microwave signal being modulated onto the corresponding comb line. Measured tap weights (*i.e.*, peak intensities of the impulse response) are obtained from the PD output recorded by the oscilloscope and are then subtracted from the designed tap weights to generate error signals, which is used to calibrate the attenuation of comb line intensity in the OSS. **Fig. 4(b)** shows the flowchart of the impulse response reshaping process in the feedback control loop at each iteration, where $b$ and $b'$ are the measured impulse response and calibrated tap coefficients, respectively. When the calibrated data $b'$ is returned to the OSS, one iteration is completed. Afterwards, the reshaped comb lines after calibration are treated as new data $b$ and used as new input to the feedback loop for subsequent iterative processes. The system can effectively reduce the error caused by the non-ideal impulse response of the system after several iterations, thus making the output impulse response approach the ideal impulse response.

For the transversal filter system in **Fig. 1(a)**, the tap coefficients $a_n$ ($n = 0, 1, 2, …, N–1$) can be either positive or negative values. In the context of the microcomb-based MWP transversal filter systems, the different signs of the tap coefficients can be realized by dividing all the wavelength channels into two groups (one with positive tap coefficients and the other with negative tap coefficients) and introducing a phase difference of $\pi$ between them. To introduce the $\pi$ phase difference, one can either use a dual-drive Mach-Zehnder modulator (DD-MZM) that has two complementary output ports to replace the EOM in **Fig. 3(a)**, or employ the complementary output ports of the OSS connecting to a balanced PD (BPD) as shown in **Fig. 4(a)**. The former needs separate spectral reshaping processes for the two groups of wavelength channels, which would increase the complexity of the feedback control. Whereas, this is not necessary for the latter because the two groups of wavelength channels are combined for differential detection and only generate one output microwave signal for impulse response reshaping. As a result, impulse response reshaping method shows advantages in achieving a low system complexity compared to the spectral power reshaping approach for processing functions in which both positive and negative tap coefficients are needed.

### B. Two-stage feedback control

On the basis of one-stage feedback control in subsection A, two-stage feedback control with two feedback loops can be used to further improve the comb shaping accuracy in microcomb-based MWP transversal filter systems.

**Fig. 5** presents a schematic diagram of two-stage feedback control based on spectral power reshaping. The spectrum of the initially generated microcomb is pre-shaped via the first OSS. Subsequently, the second OSS is employed to further shape the pre-shaped microcomb based on designed tap coefficients. The first-stage feedback control before the transversal filter module is used to calibrate the intensity errors of comb lines at the comb pre-shaping stage, and the second-stage feedback control, which is implemented within the transversal filter module, focuses on addressing the intensity errors of tap coefficients. The two-stage feedback control based on spectral power reshaping includes all the optical components of the microcomb-based MWP transversal filter system in the feedback loops. This allows for the mitigation of comb shaping inaccuracies induced by all different optical components.

By incorporating the process flow shown in **Fig. 4(b)**, the flowchart of the two-stage feedback control based on spectral power reshaping is shown in **Fig. 6**. First, spectral power reshaping is used to calibrate the pre-shaped microcomb, resulting in the generation of a microcomb with uniform comb lines that is then directed to the transversal filter module. In the transversal filter module, spectral power reshaping is also used to correct the error of tap coefficients, yielding the ultimate output following the two-stage reshaping and feedback control.

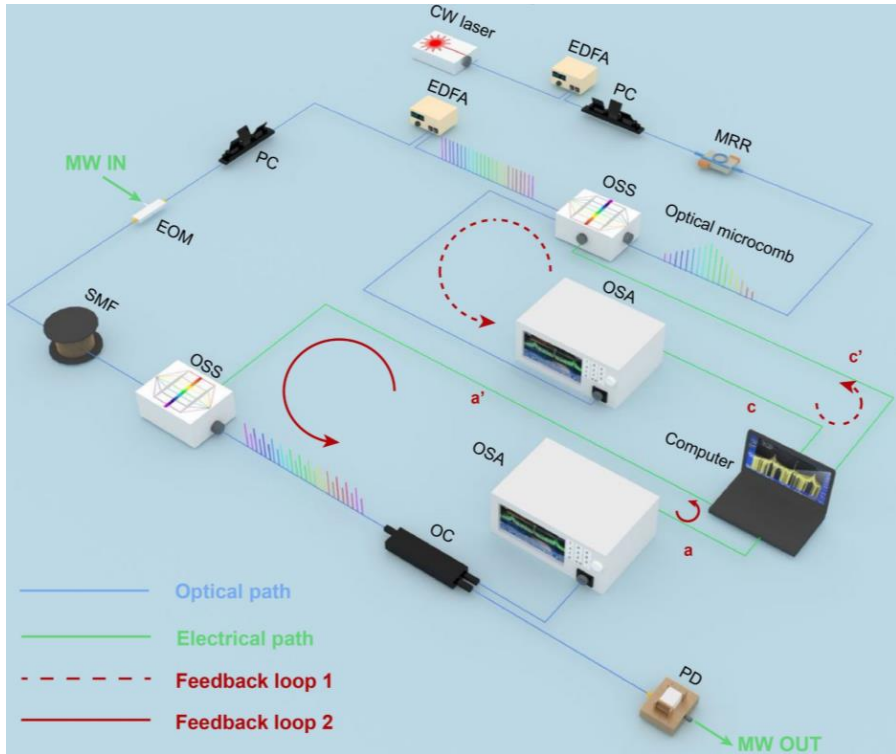

**Fig. 5.** Feedback control based on two-stage spectral power reshaping in a microcomb-based MWP transversal filter system. CW laser: continuous-wave laser. EDFA: erbium-doped fiber amplifier. PC: polarization controller. MRR: micro-ring resonator. OSS: optical spectral shaper. OSA: optical spectrum analyzer. EOM: electro-optic modulator. SMF: single-mode fiber. OC: optical coupler. PD: photo detector.

**Fig. 7** shows a schematic diagram of two-stage feedback control based on synergic spectral power reshaping and impulse response reshaping. The first OSS is employed to pre-shape the spectrum of the initially generated microcomb, and the pre-shaped comb lines are then shaped according to the designed tap coefficients via the second OSS. The first-stage feedback control based on spectral power reshaping before the transversal filter module is used to calibrate the intensity errors of comb lines at the comb pre-shaping stage, whereas the second-stage feedback control based on impulse response reshaping within the transversal filter module can further compensate the non-ideal impulse response of the system induced by the components in the transversal filter module. By using such two-stage feedback control, all the components of the microcomb-based MWP transversal filter system are included in the feedback control loops, thus allowing for compensation of the comb shaping inaccuracy induced by different components and hence significantly reduced overall comb shaping errors.

**Fig. 8** shows the flowchart of the two-stage feedback control in **Fig. 7**, which includes the process flow in **Fig. 4(b)**. Spectral power reshaping is employed to calibrate the pre-shaped microcomb, which results in the generation of a flattened microcomb that is sent to the transversal filter module. In the transversal filter module, impulse response reshaping is used to compensate the non-ideal impulse response of the system,

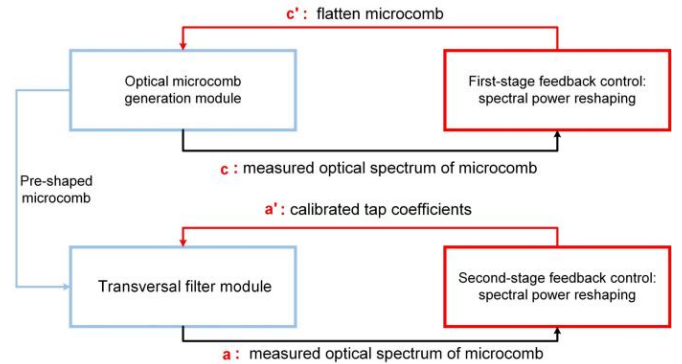

**Fig. 6.** Flowchart of two-stage feedback control based on spectral power reshaping.

which generates the ultimate output after two-stage reshaping and feedback control.

### C. Comparison of different feedback control methods

In this section, we briefly compare the feedback control methods mentioned in subsections A and B. As shown in **Table I**, although one-stage feedback control methods involve less numbers of feedback control loops and OSSs, they exhibit lower efficacy when compared to the two-stage feedback control methods. For example, employing a two-stage feedback control with an additional OSS to pre-shape the initially generated optical microcomb, which may have non-uniform power distributions, can improve the signal-to-noise ratios of the optical microcomb after amplification and expand the variation range for the tap coefficients. For spectral power reshaping method, since the OSS is inserted before the PD, it cannot compensate for the errors caused by the PD. Nevertheless, it remains effective in compensating for the errors arising from the optical microcomb, EOM, SMF, and OSS. In contrast, the impulse response reshaping method has the capability to address the errors introduced by the PD as well. For processing functions that require both positive and negative tap coefficients, the impulse response reshaping method shows advantages due to its lower system complexity. In contrast, the spectral power reshaping method needs to separate spectral reshaping processes for two sets of wavelength channels with different signs.

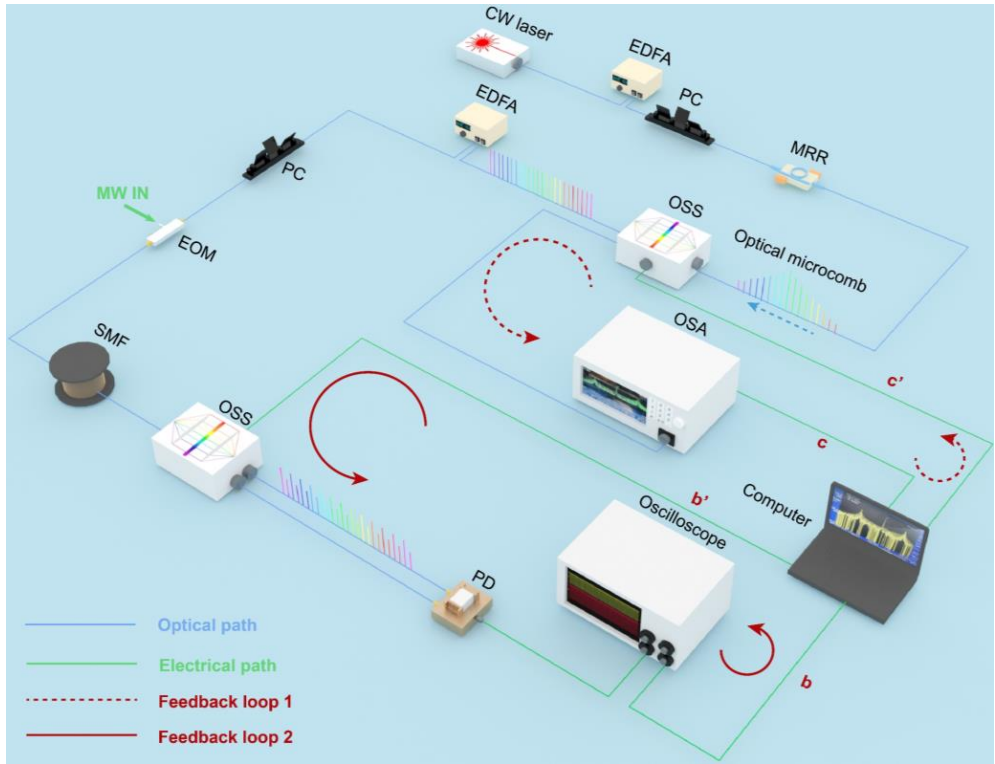

**Fig. 7.** Feedback control based on two-stage synergic spectral power reshaping and impulse response reshaping in a microcomb-based MWP transversal filter system. CW laser: continuous-wave laser. EDFA: erbium-doped fiber amplifier. PC: polarization controller. MRR: micro-ring resonator. OSS: optical spectral shaper. OSA: optical spectrum analyzer. EOM: electro-optic modulator. SMF: single-mode fiber. PD: photo detector.

## IV. EXPERIMENT RESULTS AND PERFORMANCE COMPARISON

In this section, we experimentally implement feedback control based on the different methods discussed in **Section III** and compare their performance in improving the accuracy of microcomb-based MWP transversal filter systems. To simplify, we designate the four feedback control methods in our following discussion in this section as follows: (A) one-stage spectral power reshaping, (B) one-stage impulse response reshaping, (C) two-stage spectral power reshaping, and (D) two-stage synergic spectral power reshaping and impulse response reshaping.

In our experimental demonstration, the optical microcomb was generated by a MRR made from high-index doped silica glass [44]. The doped silica offers attractive optical properties for generating optical microcombs, such as ultra-low linear loss (~0.06 dB/cm), a moderate nonlinear parameter (~233 $W^{-1} \cdot km^{-1}$), and a negligible nonlinear loss even at extremely high intensities (~25 GW $\cdot$ $cm^{-2}$) [44]. The MRR was designed to have a radius of ~592 μm, which resulted in a comb spacing of $\Delta\lambda$ = ~0.4 nm or ~49 GHz. The Q factor of the MRR was ~1.5 million. In the microcomb generation module, a CW light, amplified to ~32.1 dBm using an EDFA, was employed to pump the MRR. The polarization of the CW pump was aligned with a TE-polarized resonance of the MRR at ~1551.23 nm via a PC. When the wavelength of the CW pump was swept across the MRR's resonance, optical parametric oscillation occurred, leading to the generation of a soliton crystal optical microcomb based on thermal mode locking [31]. By using a temperature controller to maintain the MRR's temperature, we were able to sustain the stable mode-locking for the soliton crystal microcomb for a long time in our experiments (*e.g.*, ~48 hours [45]).

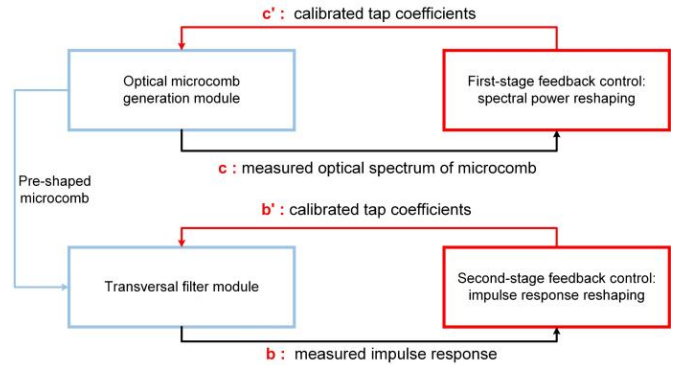

**Fig. 8.** Flowchart of two-stage feedback control based on synergic spectral power reshaping and impulse response reshaping.

TABLE I. COMPARISON OF DIFFERENT FEEDBACK CONTROL METHODS. FCL: FEEDBACK CONTROL LOOP. EOM: ELECTRO-OPTIC MODULATOR. SMF: SINGLE-MODE FIBRE. OSS: OPTICAL SPECTRAL SHAPER. PD: PHOTODETECTOR. OSA: OPTICAL SPECTRUM ANALYZER. OSC: OSCILLOSCOPE.

| Method | One-stage feedback control | | Two-stage feedback control | |
| --- | --- | --- | --- | --- |
| | Spectral power reshaping | Impulse response reshaping | Spectral power reshaping | Spectral power & impulse response reshaping |
| No. of FCLs | 1 | 1 | 2 | 2 |
| No. of OSSs | 1 | 1 | 2 | 2 |
| Included components | microcomb, EOM, SMF, OSS | microcomb, EOM, SMF, OSS, PD | microcomb, EOM, SMF, OSSs | microcomb, EOM, SMF, OSSs, PD |
| Monitoring instruments | OSA, computer | OSC, computer | OSA, computer | OSA, OSC, computer |
| Variation range for tap coefficients | small | small | large | large |
| Complexity in achieving tap coefficients with different signs | high | low | high | low |

In the transversal filter module, the comb lines were modulated by the input microwave signal via an EOM (iXblue) with an operation bandwidth of 40 GHz. The generated microwave replicas were transmitted through a spool of SMF with a dispersion parameter of $D_2$ = ~17.4 ps/nm/km and a length of $L$ = ~5.124 km, which introduced a time delay of $\Delta T = D_2 \cdot L \cdot \Delta\lambda$ = ~35.7 ps between adjacent wavelength channels. For all the methods, the comb lines were spectrally shaped by an OSS (Finisar) before the BPD to achieve the designed tap coefficients. The shaped output from the OSS was then monitored by an OSA (Anritsu) and transmitted to a computer to derive calibrated tap coefficients. For Methods C and D, the optical microcomb from the microcomb generation module was first shaped by an OSS (Finisar) to achieve uniform power distribution in the comb lines. For Methods B and D, peak intensities of the impulse response (*i.e.*, measured tap coefficients) were extracted from the output recorded by a high bandwidth real-time oscilloscope (Keysight) connected to a BPD (Finisar).

In our experimental demonstration, we first took equal tap weights (*i.e.*, $a_n$ = 1 ($n$ = 0, 1, 2, …, $N$–1)) with uniform power distribution in different taps as an example to show the effectiveness for different feedback control methods. **Fig. 9** shows the optical spectra and impulse response of selected comb lines after one-stage feedback control based on Methods A and B, respectively. In **Figs. 9(a) – (c)**, we show the results for 10, 20, and 80 comb lines, respectively. In each of them, we show the results after 2 and 10 calibration iterations. For comparison, the corresponding results without any feedback control and the power distribution before spectral shaping are also shown. As can be seen, the introducing of feedback control can effectively improve the uniformity for all different tap numbers, and the uniformity further improves as the number of iteration increases from 2 to 10.

To quantitatively characterize the uniformity of different taps in **Fig. 9**, we introduce the concept of average deviation (AD), which is defined as

$$AD = \frac{1}{N}\sum_{n=0}^{N-1}\frac{|P_n - P_{avg}|}{P_{avg}} \qquad (2)$$

where $P_n$ ($n$ = 0, 1, 2, …, $N$–1) is the power of the tap corresponding to the tap coefficient $a_i$, and $P_{avg}$ = ($P_0 + P_1 + P_2 + … + P_{n-1}$) / $N$ is the average of the powers across all the different taps. **Figs. 9(d-i)** and **(d-ii)** show the ADs versus the number of iterations for Methods A and B, respectively. For each method, we show the results for 10, 20, and 80 comb lines. The values at iteration = 0 correspond to the results without any feedback control. For both methods, the ADs decrease with increasing number of iterations, showing agreement with the trend in **Fig. 9(a) – (c)**. As the number of iterations increases, the decrease in ADs becomes more gradual, with only a minimal decrease in the ADs when the number of iterations exceeds 4. This suggests that a large number of iterations is not necessary. We also note that achieving the same AD value for more taps requires a higher number of iterations. Depite this, for all scenarios, low AD values < 0.005 can be attained after just 4 iterations.

In **Fig. 10**, we show the optical spectra and impulse response of selected comb lines after two-stage feedback control based on Methods C and D, respectively. **Figs. 10(a)** and **(b)** show the results for 10 and 20 comb lines, respectively. In each of them, we present the results after the first-stage and second-stage feedback control, and the number of iterations is 4. The power distribution of the comb lines before spectral shaping is also shown for comparison. For both methods, by employing the second-stage feedback control, the uniformity of the different taps is further improved on the basis of the first-stage feedback control. **Fig. 10(c)** shows the ADs versus the number of iterations for Methods C and D, which were calculated based on **Eq. (2)** using the results in **Figs. 10(a)** and **(b)**. The values at iteration = 0 represent the results without any feedback control. Similar to the trend observed in **Fig. 9d**, the ADs decrease with increasing number of iterations. We also note that for both methods low AD values < 0.005 can be achieved after just 2 iterations.

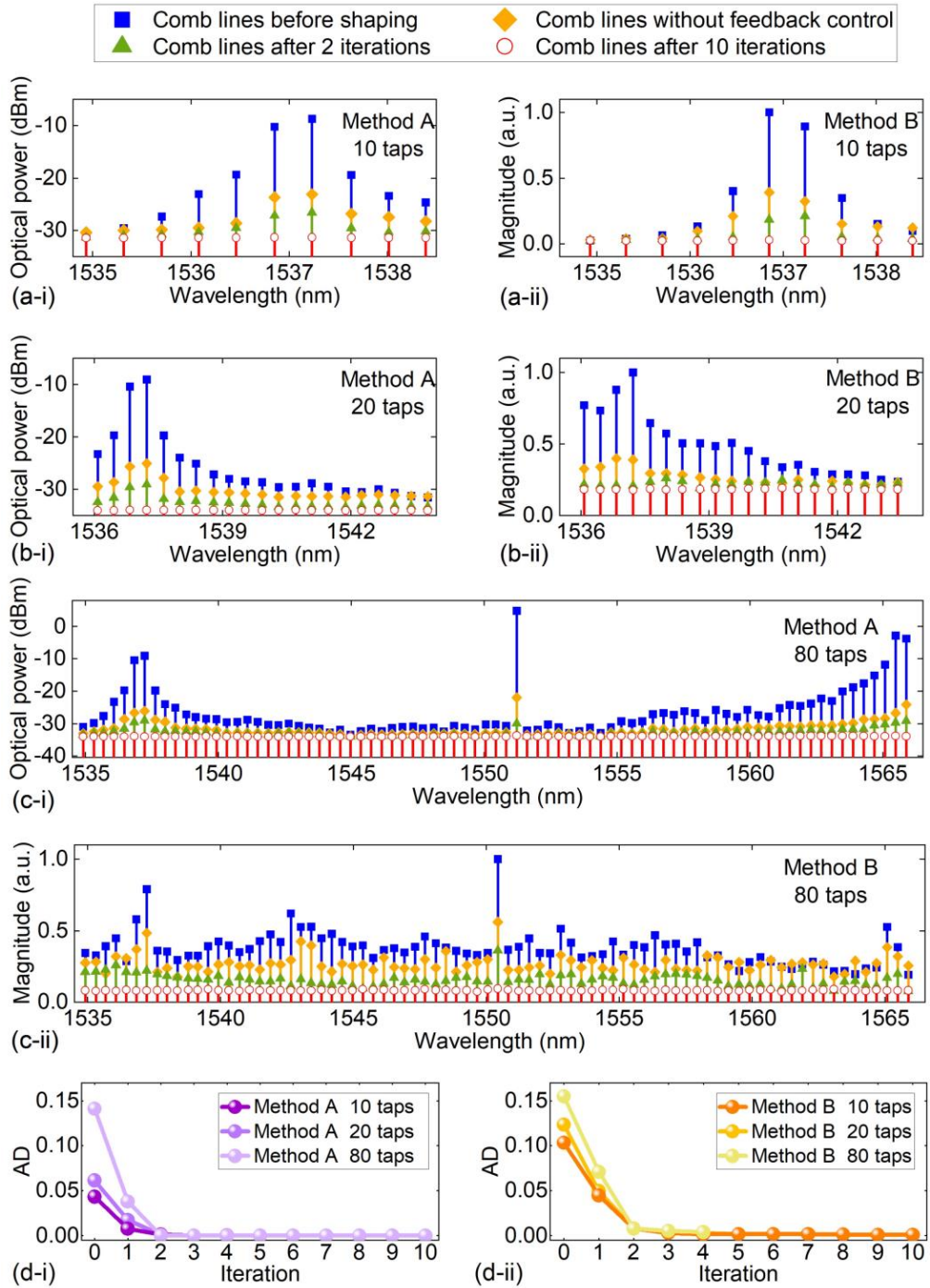

**Fig. 9.** Experimental results of one-stage feedback control. (a) – (c) Results of 10, 20, and 80 comb lines, respectively, where (i) and (ii) show the optical spectra for feedback control based on Method A and the impulse response for feedback control based on Method B, respectively. In each figure, the power distribution of the comb lines before shaping, the shaping results without any feedback control, and the shaping results after 2 and 10 calibration iterations are shown for comparison. (d) Average deviations (ADs) versus the number of iterations for (i) Method A and (ii) Method B calculated based on the results in (a) – (c).

The AD values in **Fig. 9(d-i)** were calculated based on the measured optical spectra of comb lines (which were recorded by an OSA). On the other hand, the AD values in **Fig. 9(d-ii)** were calculated based on the measured impulse responses (which were recorded by an oscilloscope). Therefore, the AD values in these two figures cannot be directly compared to evaluate the performance for Methods A and B. Similarly, the AD values in **Fig. 10(c-i)** and **Fig. 10(c-ii)** cannot be directly compared to evaluate the performance for Methods C and D. To directly compare the performance of different feedback control methods, we further employed the microcomb-based MWP transversal filter systems to perform temporal signal processing and spectral filtering. The root mean square error (RMSE) is introduced to characterize the discrepancy from the measured result to the ideal result, which can be expressed as

$$\text{RMSE} = \sqrt{\sum_{i=1}^{n} \frac{(Y_i - y_i)^2}{n}} \qquad (3)$$

where $Y_1$, $Y_2$, …, $Y_n$ are the values of ideal output, $y_1$, $y_2$, …, $y_n$ are values of measured output, and $n$ is the number of sampled points.

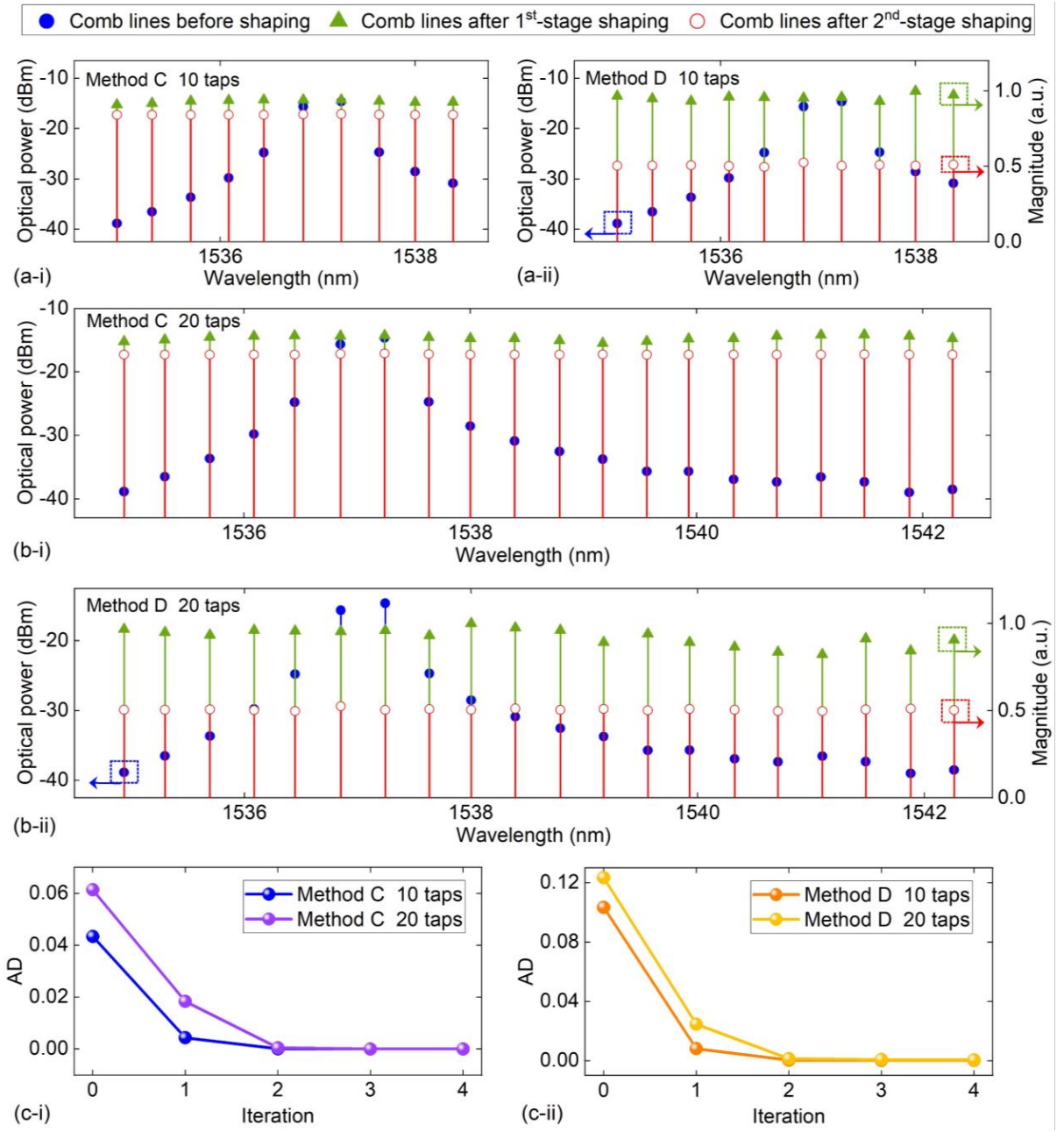

**Fig. 10.** Experimental results of two-stage feedback control. (a) – (b) Results of 10 and 20 comb lines, respectively. (i) shows the optical spectra for feedback control based on Method C. (ii) shows the optical spectra and impulse response for feedback control based on Method D. In each figure, the power distribution of the comb lines before shaping and the shaping results after the 1$^{st}$-stage and the 2$^{nd}$-stage feedback control are shown for comparison. (c) Average deviations (ADs) versus the number of iterations for (i) Methods C and (ii) Method D calculated based on the results in (a) and (b).

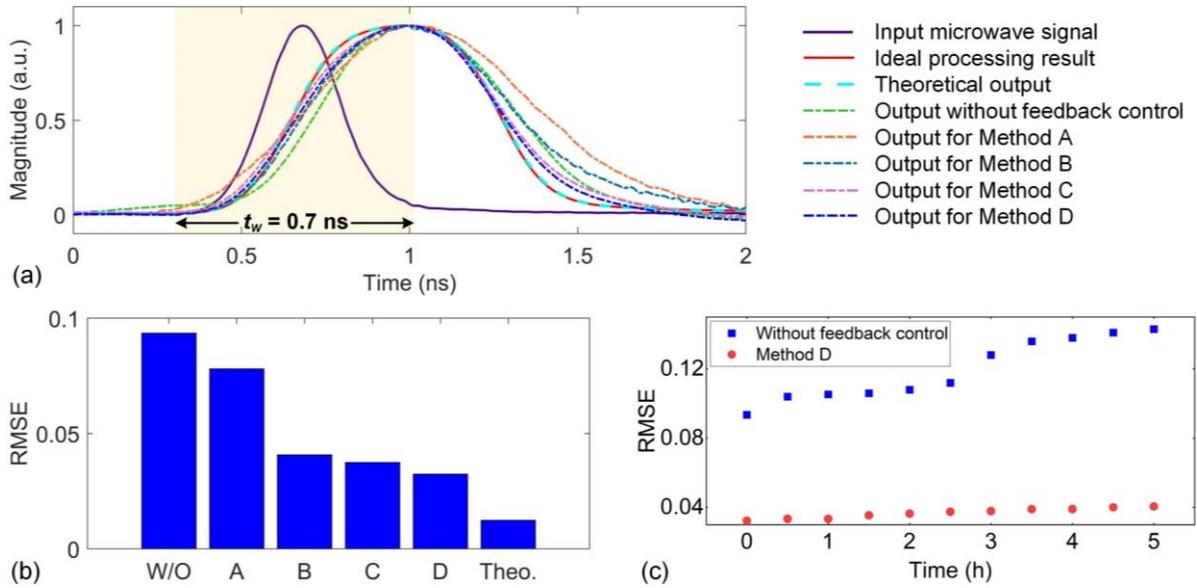

**Fig. 11.** Performance comparison of different feedback control methods for microcomb-based MWP transversal filter system that performs temporal integration. (a) Measured output waveforms when employing feedback control methods A – D. For comparison, the waveforms for the input Gaussian pulse, the ideal integration result without any errors, the theoretical output assuming that all the system components have perfect responses, and the system output without feedback control are also shown. The shaded area represents the integration window of $t_w = 0.7$ ns. (b) Root mean square errors (RMSEs) between the ideal integration result and the other results in (a). The results for the theoretical system output assuming that all the system components have perfect responses and the integration result without feedback control are denoted as 'Theo.' and 'W/O', respectively. (c) RMSEs versus system running time for the system without feedback control and with feedback control based on Method D.

In **Fig. 11**, we compare the performance for the microcomb-based MWP transversal filter system that performs temporal integration. A Gaussian pulse with a full width at half maximum (FWHM) of ~0.2 ns generated by an AWG (Keysight) was used as the input microwave signal. The tap number was $N = 20$, which resulted in an integration window of $t_w = N \times \Delta T = ~0.7$ ns. **Fig. 11(a)** shows the temporal waveform for the input Gaussian pulse and the measured output waveforms when employing feedback control methods A – D. For comparison, the ideal integration result without any errors, the system output without feedback control, and the theoretical output assuming that all the system components have perfect responses are also shown. In our calculation of the ideal integration result and the theoretical output, we used the output waveform from the AWG captured by an oscilloscope as the input signal, which helped circumvent additional errors induced by the difference between the practical input pulse and the ideal Gaussian pulse. As can be seen, the implementation of feedback control results in a closer match between the system output and the ideal output compared to the system without any feedback control. Among all the different feedback control methods, the output of Method D exhibits the closest match with the ideal integration result, highlighting its superiority in improving the processing accuracy.

**Fig. 11(b)** shows the RMSEs between the ideal output and the other results in **Fig. 11(a)**. The RMSE for the theoretical output assuming that all the system components have perfect responses is induced by the theoretical limitation of the transversal filter system, which arises from the theoretical approximation of a continuous impulse response (that corresponds to an infinite tap number) using a practical system having a finite tap number. The RMSE values for Methods A and B are higher than those for Methods C and D, showing a trend similar to the results in **Figs. (9)** and **(10)** and further confirming the improved performance for the two-stage feedback control. Although Method D exhibits the lowest RMSE value among the four methods, it is still higher than the RMSE for the theoretical output, indicating that there were remaining errors in the system that cannot be compensated by employing the current feedback control methods. These errors were mainly fast varying errors, such as the amplitude and phase errors induced by microcombs and the PD.

To assess the system's stability in operating for a long time, we measured the system outputs both with the implementation of feedback control based on Method D and without feedback control. The outputs were recorded every 30 minutes, spanning a period of 5 hours. **Fig. 11(c)** shows the calculated RMSEs based on the recorded system outputs versus running time. As can be seen, the absence of feedback control resulted in an obvious decline in the processing accuracy as the system ran over time. In contrast, the implementation of feedback control enabled the system to maintain stable operation with a high level of processing accuracy over a significantly extended duration. These results further highlight the importance of implementing feedback control in practical systems.

We also compare the performance for the microcomb-based MWP transversal filter system that performs low-pass filtering. **Fig. 12(a)** shows the measured RF response of the system recorded by a vector network analyzer (VNA, Anritsu) when implementing feedback control based on Methods A – D. The RF response for the system without feedback control and the theoretical RF response calculated based on **Eq. (1)** assuming that all the system components have perfect response are also shown for comparison. As expected, incorporating feedback control leads to a more accurate alignment between the response of the practical system and the theoretical response. Similar to that in **Fig. 11(a)**, the system response when employing feedback control based on Method D exhibits the lowest discrepancies in comparison to the theoretical response. **Fig. 12(b)** shows the calculated RMSEs between the theoretical RF response and the measured RF response based on the results in **Fig. 12(a)**. **Fig. 12(c)** compares the RMSEs versus running time for the systems with feedback control based on Method D and without feedback control. The results in these figures show similar trends as those in **Figs. 11(b)** and **(c)**, which confirms that the implementation of feedback control works for not only temporal processing but also spectral filtering.

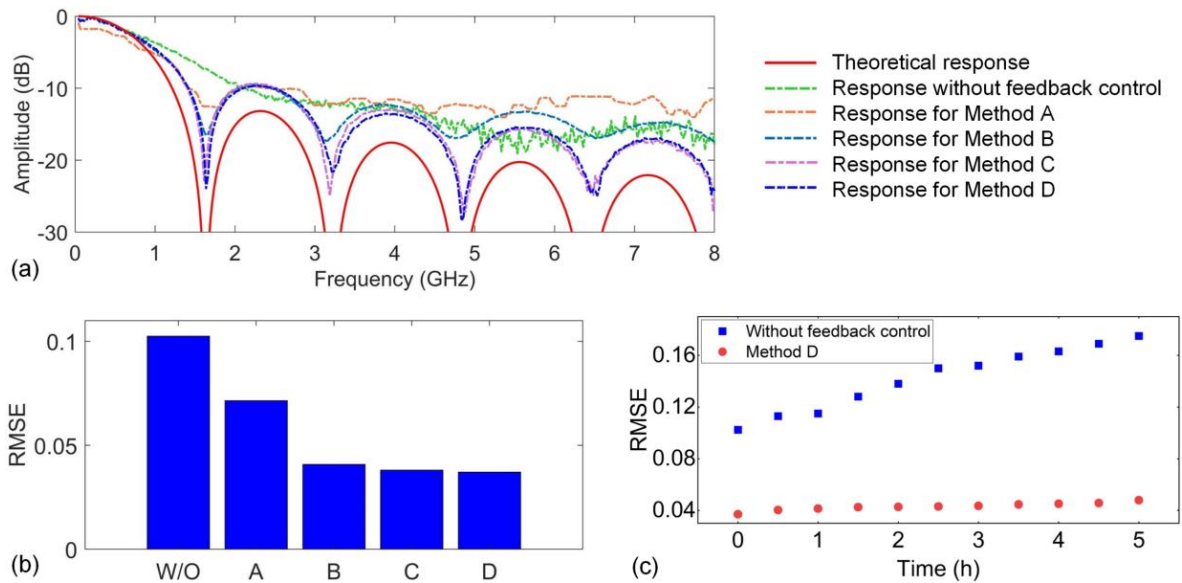

**Fig. 12.** Performance comparison of different feedback control methods for microcomb-based MWP transversal filter system that performs low-pass filtering. (a) Measured RF response when employing feedback control Methods A – D. For comparison, the theoretical RF response assuming that all the system components have perfect responses and the RF response without feedback control are also shown. (b) RMSEs between the theoretical RF response and the other results in (a). The result for the system without feedback control is denoted as 'W/O'. (c) RMSEs versus system running time for the system without feedback control and with feedback control based on Method D.

## V. DISCUSSION

Feedback control techniques play an important role across various applications due to their capability to improve both system accuracy and stability. As evidenced by our results in previous sections, feedback control in microcomb-based MWP transversal filter systems can effectively reduce errors caused by experimental components and enhance the system stability in working for long times. Despite this, though, there are still challenges that need to be addressed. In this section, we discuss the limitations and future prospects for further improving the feedback control performance.

Achieving stable mode-locking of optical microcombs is critical for their practical applications, including their use in MWP transversal filter systems. In our previous discussion, we assumed that the generated microcomb was stable and did not account for the errors induced by the microcomb's instability in the feedback control, while this needs to be considered for application scenarios that require the system to operate for long time periods. In the past decade, many approaches have been proposed to achieve stable mode-locking of microcombs, such as frequency scanning [45], power kicking [46], forward and backward tuning [47], two-colour pumping [48], EO modulation [49], self-injection locking [50], filter-driven FWM [51], integrated heaters [52], and self-referencing [53], cryogenic cooling [54], auxiliary laser heating [55], and nonlinear dynamics engineering [56]. Recently, significant progress has also been achieved in turnkey soliton microcomb generation [56] and piezoelectric feedback control of microcombs via integrated actuators [57]. These methods open up new avenues towards achieving mode-locking of microcombs with high stability in practical applications.

The performance of feedback control in microcomb-based MWP transversal filter systems is also affected by the monitoring instruments such as the OSA and oscilloscope. To achieve precise spectral power reshaping, a high sampling resolution of the OSA is needed for detailed capture of fine optical spectral characteristics. On the other hand, the high sampling resolution also results in a large number of sampling points, which increases the processing time and hence the overall feedback control time. As a result, there is a trade-off that needs to be considered when selecting the resolution of OSA in practical applications. For impulse response reshaping, utilizing the oscilloscope's averaging function, which involves capturing waveforms for multiple times and then averaging them point by point, can be effective in minimizing the influence of noises and improving the accuracy of feedback control. Nevertheless, this could also result in the increase of processing time and the overall in Section III can also be used for realizing feedback control in feedback control time, which introduces another trade-off in the selection of the averaging times of oscilloscope.

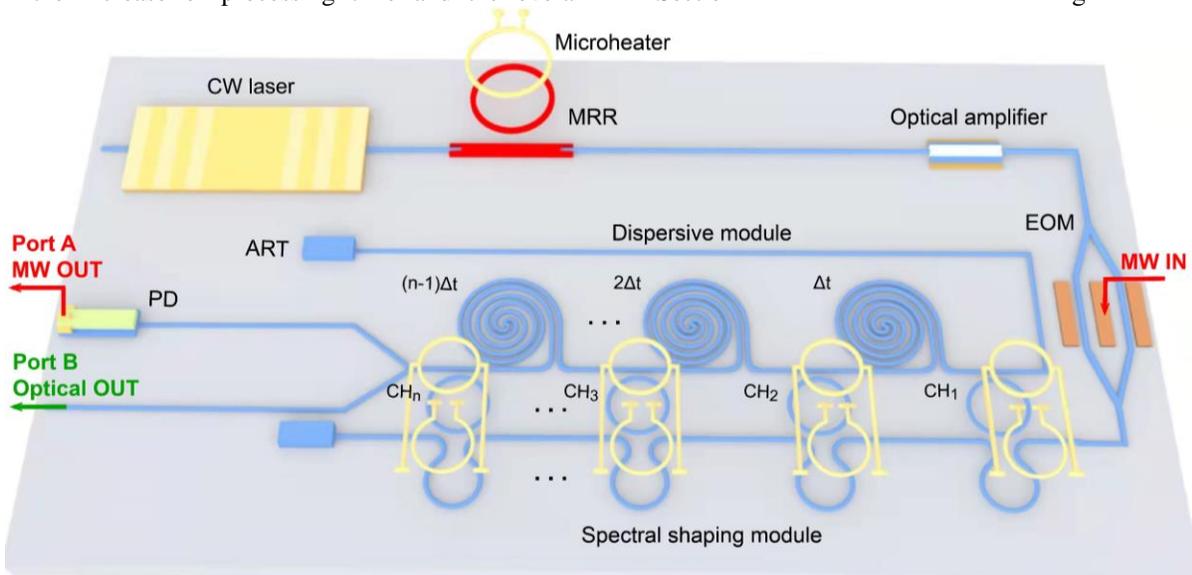

**Figure 13.** A schematic showing the concept of a monolithically integrated microcomb-based MWP transversal filter system. CW laser: continuous-wave laser. MRR: microring resonator. EOM: electro-optic Mach-Zehnder modulator. PD: photodetector. ART: anti-reflection termination.

Recently, there has been significant progress towards monolithically integrated microcomb-based MWP transversal filter systems [14]. Although employing integrated microcomb sources to replace discrete laser arrays already brings many benefits in terms of SWaP, cost, and complexity, there is much more to be gained by further increasing the integration level of the overall system. **Fig. 13** shows the schematic of an on-chip microcomb-based MWP transversal filter system, where the CW laser, optical amplifier, EO modulator, dispersive module, optical spectral shaping module, and PD are all implemented in their integrated forms. In principle, all the components in the microcomb-based MWP transversal filter system can be integrated on the same chip, and on-chip CW lasers [58], optical amplifiers [59], EO modulators [60], dispersive elements [61], optical spectral shapers [62, 63], and PDs [64] have all been demonstrated. On the basis of these integrated components, some complicated subsystems such as microcomb generation module consisting of heterogeneously integrated pump lasers and microresonators [65] and spectral shaping arrays [66, 67] have also been realized.

For on-chip microcomb-based MWP transversal filter systems, feedback control is also needed to improve the shaping accuracy and system stability, although the development of feedback control technologies in these systems is still in its nascent stages. Except for the previously mentioned methods for achieving stable mode-locking in the microcomb generation module, the transversal filter module also needs feedback control to compensate the comb shaping errors induced by different components such as EO modulator, dispersive module, spectral shaping module, and PD. The spectral power reshaping and impulse response reshaping methods discussed on-chip microcomb-based MWP transversal filter systems. For spectral power reshaping, the feedback control can be realized by comparing the output optical signal at Port B in **Fig. 13** with the ideal one to generate the calibrated signal that is applied to the microheaters in the spectral shaping module to tailor the intensity of comb line in each channel. For impulse response reshaping, channel-by-channel power reshaping can be realized by modulating the same input RF signal onto the corresponding comb line and comparing the measured RF output signal at Port A in **Fig. 13** with the designed tap weights to generate the calibrated signal for adjusting the microheaters. Recently, a self-calibrating photonic integrated circuits has been demonstrated [68], where the impulse response calibration was realized by incorporating an optical reference path to establish an on-chip Kramers-Kronig relationship and then employing a fast-converging algorithm to calculate the tap-value errors from the measured and desired impulse responses. This offers new possibilities for realizing stable and accurate feedback control in on-chip microcomb [80-103] based microwave photonic transversal filter systems, [104-136] potentially aided in terms of integration with the advanced functionalities offered by 2D materials [137-161]. These devices will benefit from the advances made in microcombs generally and microwave applications of microcombs. [162-223]

## VI. CONCLUSON

In summary, we demonstrate the effectiveness of introducing of feedback control in microcomb-based MWP transversal filter systems to improve their performance. We propose and experimentally demonstrate four different feedback control methods. The experimental results show that the implementation of feedback control enhances not only the accuracy of comb line shaping, temporal signal processing, and spectral filtering, but also the system stability in working for long times. In addition, the two-stage synergic spectral power reshaping and impulse response reshaping exhibited the best performance in improving the system performance among the four methods. Finally, we discuss the challenges and prospects for improving the performance of feedback control in the systems implemented by both discrete components and integrated chips. These results provide a useful guide for the implementation of feedback control in microcomb-based MWP filter systems, which paves the way for improving their performance in practical applications.